\documentclass[iop]{emulateapj}
\usepackage{epstopdf}
\graphicspath{{.}{images/}}
\usepackage{array}
\usepackage{multirow}

\shorttitle{HAT-P-2's Planet-Induced Pulsations}
\shortauthors{de Wit et al.}

\begin{document}

\title{Planet-Induced Stellar Pulsations in HAT-P-2's Eccentric  System} 

\author
{Julien de Wit\altaffilmark{1}, Nikole K. Lewis\altaffilmark{2}, Heather A. Knutson\altaffilmark{3}, Jim Fuller\altaffilmark{4,5},Victoria Antoci\altaffilmark{6}, Benjamin J. Fulton\altaffilmark{7,17}, Gregory Laughlin\altaffilmark{8},  Drake Deming\altaffilmark{9}, Avi Shporer\altaffilmark{10,18}, Konstantin Batygin\altaffilmark{3}, Nicolas B. Cowan\altaffilmark{11}, Eric Agol\altaffilmark{12}, Adam S. Burrows\altaffilmark{13},   Jonathan J. Fortney\altaffilmark{14}, Jonathan Langton\altaffilmark{15} and Adam P. Showman\altaffilmark{16}}

\altaffiltext{1}{Department of Earth, Atmospheric and Planetary Sciences, MIT, 77 Massachusetts Avenue, Cambridge, MA 02139, USA}
\altaffiltext{2}{Space Telescope Science Institute, 3700 San Martin Drive, Baltimore, MD 21218, USA}
\altaffiltext{3}{Division of Geological and Planetary Sciences, California Institute of Technology, Pasadena, CA 91125, USA}
\altaffiltext{4}{TAPIR, Walter Burke Institute for Theoretical Physics, Mailcode 350-17, California Institute of Technology, Pasadena, CA 91125, USA}
\altaffiltext{5}{Kavli Institute for Theoretical Physics, University of California, Santa Barbara, CA 93106, USA}
\altaffiltext{6}{Stellar Astrophysics Centre, Department of Physics and Astronomy, Aarhus University, Ny Munkegade 120, 8000 Aarhus C, Denmark}
\altaffiltext{7}{Institute for Astronomy, University of Hawaii, Honolulu, HI 96822, USA}
\altaffiltext{8}{Department of Astronomy, Yale University, New Haven, CT 06511, USA}
\altaffiltext{9}{Department of Astronomy, University of Maryland at College Park, College Park, MD 20742, USA}
\altaffiltext{10}{Jet Propulsion Laboratory, California Institute of Technology, 4800 Oak Grove Drive, Pasadena, CA 91009, USA}
\altaffiltext{11}{Department of Physics, Department of Earth and Planetary Sciences, McGill University, 3550 rue University, Montreal, QC H3A 2A7, Canada}
\altaffiltext{12}{Department of Astronomy, University of Washington, Seattle, WA 98195, USA}
\altaffiltext{13}{Department of Astrophysical Sciences, Princeton University, Princeton, NJ 08544, USA}
\altaffiltext{14}{Department of Astronomy and Astrophysics, University of California, Santa Cruz, CA 95064, USA}
\altaffiltext{15}{Department of Physics, Principia College, Elsah, IL 62028, USA}
\altaffiltext{16}{Lunar and Planetary Laboratory, University of Arizona, Tucson, AZ 85721, USA}
\altaffiltext{17}{NSF Graduate Research Fellow}
\altaffiltext{18}{Sagan Postdoctoral Fellow}

\begin{abstract}
Extrasolar planets on eccentric short-period orbits provide a laboratory in which to study radiative and tidal interactions between a planet and its host star under extreme forcing conditions.  Studying such systems probes how the planet's atmosphere redistributes the time-varying heat flux from its host and how the host star responds to transient tidal distortion. Here, we report the insights into the planet-star interactions in HAT-P-2's eccentric planetary system gained from the analysis of $\sim$350 hours of 4.5 $\mu$m observations with the Spitzer Space Telescope. The observations show no sign of orbit-to-orbit variability nor of orbital evolution of the eccentric planetary companion, HAT-P-2\,b. The extensive coverage allows us to better differentiate instrumental systematics from the transient heating of HAT-P-2\,b's 4.5 $\mu$m photosphere and yields the detection of stellar pulsations with an amplitude of approximately 40 ppm. These pulsation modes correspond to exact harmonics of the planet's orbital frequency, indicative of a tidal origin. Transient tidal effects can excite pulsation modes in the envelope of a star but, to date, such pulsations had only been detected in highly-eccentric stellar binaries. Current stellar models are unable to reproduce HAT-P-2's pulsations, suggesting that our understanding of the interactions at play in this system is incomplete. 
\end{abstract}


\section{Introduction}

Owing to its large mass (8 $M_{\rm Jup}$), short orbital period ($P\sim$5.63\,d), and eccentricity ($e\sim0.5$), the hot Jupiter HAT-P-2\,b \citep{Bakos2007} is a favored target for the study of planet-star interactions \citep{Fortney2008,Jordan2008,Langton2008,Fabrycky2009,Levrard2009,Hartman2010,Cowan2011,Cebron2013,Kataria2013,Lewis2013,Lewis2014,Salz2016}. Measurements of the flux variations caused by its transient heating obtained at 3.6, 4.5, and 8.0 $\mu$m with \textit{Spitzer} yielded the first insights into the planet's response to transient heating \citep{Lewis2013} but were not reproducible by atmospheric models \citep{Lewis2014}, in particular at 4.5 $\mu$m. Here, we present the analysis of all the observations of HAT-P-2\,b obtained at 4.5 $\mu$m with \textit{Spitzer}---including two new primary and sixteen new secondary eclipses---whose extensive coverage allows us to better differentiate instrumental systematics from the transient heating of HAT-P-2\,b's 4.5 $\mu$m photosphere and yields the detection of unexpected pulsations with an amplitude of 40 ppm.

\section{Observations \& Analysis}
\label{sec:obs}

The observations total 350 hours (2.9 million 32x32-pixels subarray images with a 0.4 sec exposure time) and include (1) a full-orbit phase curve obtained between 2011 July 9 and 2011 July 15 \citep{Lewis2013}, (2) sixteen new secondary eclipses---including two partial phase curves each covering 30 hrs post periastron passage---obtained between April 2013 and October 2013 (PI: H. Knutson), and (3) two new primary and two new secondary eclipses obtained between October 2015 and November 2015 (PI: N. Lewis).

We extract the photometry following the method detailed in \citet{deWit2016}. While \citet{Lewis2013} used a fixed aperture of 2.25 pixels to minimize the scatter in the final solution, we use here for all the Astronomical Observation Requests (AORs) time-varying apertures with a radius equal to the square root of the noise pixel parameter $\tilde{\beta}$ \citep[see, e.g.,][]{Mighell2005,Lewis2013} with an AOR-dependent constant offset estimated to minimize the relative amount of red noise in the final time series of each AOR (see \textbf{Online Table}). The resulting aperture radius has an average across all AORs of 1.75 pixels. We trim outliers from the resulting time series for each visit, discarding 0.8\% of the images overall.

As for the photometry extraction, we follow the standard procedures described in \citet{deWit2016} to analyze the photometry, allowing us to reach photometric precisions of 75 ppm per 1-hr bin. We update the combined instrumental and astrophysical model to include functional forms to account for the transient heating of the planet's atmosphere due to its eccentric orbit and the stellar pulsations. The resulting phase-folded photometry is shown in Figure\,\ref{fig:photometry}. 

\subsection{Nominal Model}

We model transits and occultations following  \citet{Mandel2002} and allow the eclipse times to vary to search for possible orbital evolution. We model HAT-P-2's transient heating using the functional forms introduced in \citet{Lewis2013}. We find that the asymmetric Lorentzian function is favored over the function based on harmonics in orbital phase ($\Delta BIC$ = -7). This difference is due to two new partial phase curves obtained after periastron passage together with sixteen new occultations that provide a stronger leverage on HAT-P-2\,b's flux modulation around occultation allowing a better disentanglement from the instrumental systematics.

\subsection{Pixelation Effect and Intrapixel Sensitivity Variations}

The change in position of the target's point-spread function (PSF) over a detector with non-uniform intrapixel sensitivity leads to apparent flux variations that are strongly correlated with the PSF position. We account for this effect using the same implementation of the pixel-mapping method as in \citet{Lewis2013}---see, e.g., \citet{Wong2015} and \citet{Ingalls2016} for detailed method comparisons. 

\subsection{Detector Ramp}

IRAC observations often display an exponential increase in flux at the beginning of a new observation. The standard procedure for 4.5 $\mu$m observations is to trim the first hour at the start of each observation and subsequent downlinks which reduces the complexity of fits with minimal loss of information on the eclipse shape and time \citep{Lewis2013}. However, we find here that the photometry around the primary and secondary eclipses shows a positive trend. Individual primary and secondary eclipse fits allowing for the correction of a linear trend in the photometry show that the trend is consistent across all epochs and has a slope of $65\pm6$ ppm/hr (see \textbf{Online Table}). Similarly to \citet{Stevenson2012}, we find that this trend is best accounted for with a simple exponential function. 

\subsection{Pulsations}

While fitting HAT-P-2's photometry with the standard models introduced above, we observed additional signals in the photometry in the forms of oscillations (Figure\,\ref{fig:photometry}). We account for the oscillations by including sine functions in our global fit---details in Sections\,\ref{sec:puls} and \ref{sec:dis_puls}. 
 
\subsection{Time-Correlated Noise and Uncertainty Estimates}

We find that the standard deviation of our best-fit residuals is a factor of 1.11 higher than the predicted photon noise limit at 4.5 $\mu$m. We expect that this noise is most likely instrumental in nature, and account for it in our error estimates using a scaling factor $\beta_{red}$ for the measurement errobars \citep{Gillon2010a}. We find that the average $\beta_{\rm red}$ is 1.3, the maximum values corresponding to a timescale of $\sim$50 minutes.

\section{Results}
\label{sec:res}

\subsection{Orbit-to-Orbit Variability}

\begin{figure}
  \centering\includegraphics[trim=0mm 00mm 0mm 00mm, clip=true,width=9cm,height=!]{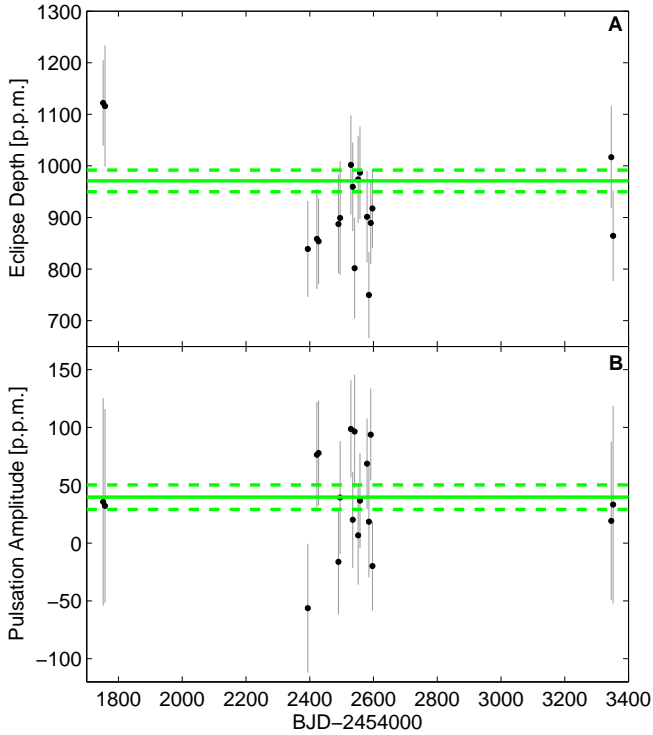}
  \vspace{-0.0cm}
  \caption{Estimates of occultation depth \textbf{(A)} and pulsation amplitude \textbf{(B)} for the eighteen occultations observed in the 4.5 $\mu$m \textit{Spitzer} band (\textbf{Online Table}). \textbf{(A)} The green lines indicate the best-fit occultation depth and corresponding 1 sigma uncertainties from the global fit. The good agreement between the individual occultation depth estimates and the global fit implies no orbit-to-orbit variability for HAT-P-2\,b's 4.5 $\mu$m photosphere and highlights the consistency of \textit{Spizter} measurements. \textbf{(B)} The individual amplitudes scatter homogeneously around 40 ppm. No single occultation can provide a significant detection of the pulsation signal due to noise limitations with a single observation.  However, all the individual occultation amplitudes are consistent with the combined pulsation signal. The green lines indicate the mean of the individual fits and their standard deviation, which is in good agreement with each individual estimate.}
  \label{fig:ind_eclipse_depths_and_oscillation}
\end{figure}

We find no sign of variability across the eighteen occultations obtained with \textit{Spitzer} at 4.5 $\mu$m. The occultation depths estimated via individual fits are shown in Figure\,\ref{fig:ind_eclipse_depths_and_oscillation}(A) and are all consistent with the global-fit estimate within $\sim 1 \sigma$. This implies that the thermal structure of HAT-P-2\,b's 4.5 $\mu$m photosphere around occultation shows no sign of orbit-to-orbit variability.
This lack of significant orbit-to-orbit variability is consistent with the predictions presented in \citet{Lewis2014}. 

The reproductibility of the parameter estimates across eighteen occultations obtained over four years also further emphasizes the reliability of current techniques used for the acquisition and analysis of Warm \textit{Spitzer} measurements \citep[see also e.g.,][]{Wong2015,Wong2015a,Ingalls2016}.

\subsection{HAT-P-2's Orbital Parameters and Ephemeris}
\label{sec:param}

We find no sign of orbital evolution from the estimated times of primary and secondary eclipses (\textbf{Online Table}). We derive a new ephemeris from a simultaneous fit of our measured primary and secondary eclipse times, those from the literature \citep[all but the 4.5~$\mu$m eclipse times gathered in][]{Lewis2013} and HAT-P-2's radial velocity (RV) measurements\footnote{RV measurements obtained by the California Planet Search (CPS) team with the HIRES instrument \citep{Vogt1994} on Keck using the CPS pipeline \citep{Howard2009}.} (\textbf{Online Table}). Our new observations extend the previous baseline by a factor of two and the number of occultations by a factor of five yielding ultra-precise estimates of HAT-P-2\,b's orbital period, midtransit time, orbital eccentricity, and argument of periastron:
\begin{equation}
\left.
\begin{array}{l c l}      
    P & = & 5.6334675\pm 1.3 \times10^{-6} \mbox{days}\\
    T_{c,0} & = & 2455288.84969\pm0.00039 \mbox{ BJD}\\
    e & = & 0.51023\pm0.00042\\
    \omega & = & 188.44\pm0.43^{\circ}\\
\end{array}.\right.
\end{equation}

Our global fit of HAT-P-2's photometry also yield improved constraints on the transit depth (4941$\pm$39 ppm), the occultation depth (971$\pm$21 ppm, which translates into an hemisphere-averaged brightness temperature of 2182$\pm$27 K), the orbital inclination (86.16$\pm$0.26$^\circ$), and stellar density (0.434$\pm$0.020 g/cm$^{3}$). We also derive from the ultra-precise eclipse times the 3-$\sigma$ upper limit on the variation of HAT-P-2\,b's orbital eccentricity and argument of periastron, respectively 0.0012 and 0.90$^{\circ}$ per year.

\subsection{HAT-P-2\,b's Transient Heating}

The extensive coverage of HAT-P-2\,b's occultation and post-periastron passage allows us to disentangle further systematics from the planetary flux modulation. The result is a phase curve (Figure\,\ref{fig:photometry}) whose main difference with \citet{Lewis2013} is a lower minimum---322$\pm$69 ppm, which translates into a hemisphere-averaged temperature of 1327$\pm$106 K. We find an excellent agreement on the timing and amplitude of the planetary flux peak, 5.40$\pm$0.57 hr after periastron passage and 1178$\pm$34 ppm---which translates into a hemisphere-averaged temperature of 2425$\pm$40 K. The timescales for the planetary flux increase and decrease are respectively 5.5$\pm$1.1 hr and 10.3$\pm$1.5 hr. Because of the orbital geometry of HAT-P-2\,b with periastron passage occurring midway between transit and occultation, the measured flux increase and decrease timescales are likely an overestimate of the true radiative timescale 
of HAT-P-2\,b's atmosphere near the 4.5~$\mu$m photosphere, but are consistent with a short atmospheric radiative timescale of a few hours.  The results presented here are in excellent agreement with the predictions from three-dimensional general circulation models presented in \citet{Lewis2014}.

\subsection{HAT-P-2's Pulsations}
\label{sec:puls}

\begin{figure*}[!ht]
  \vspace{-0.0cm}\centering\includegraphics[trim=0mm 0mm 0mm 00mm, clip=true,width=17cm,height=!]{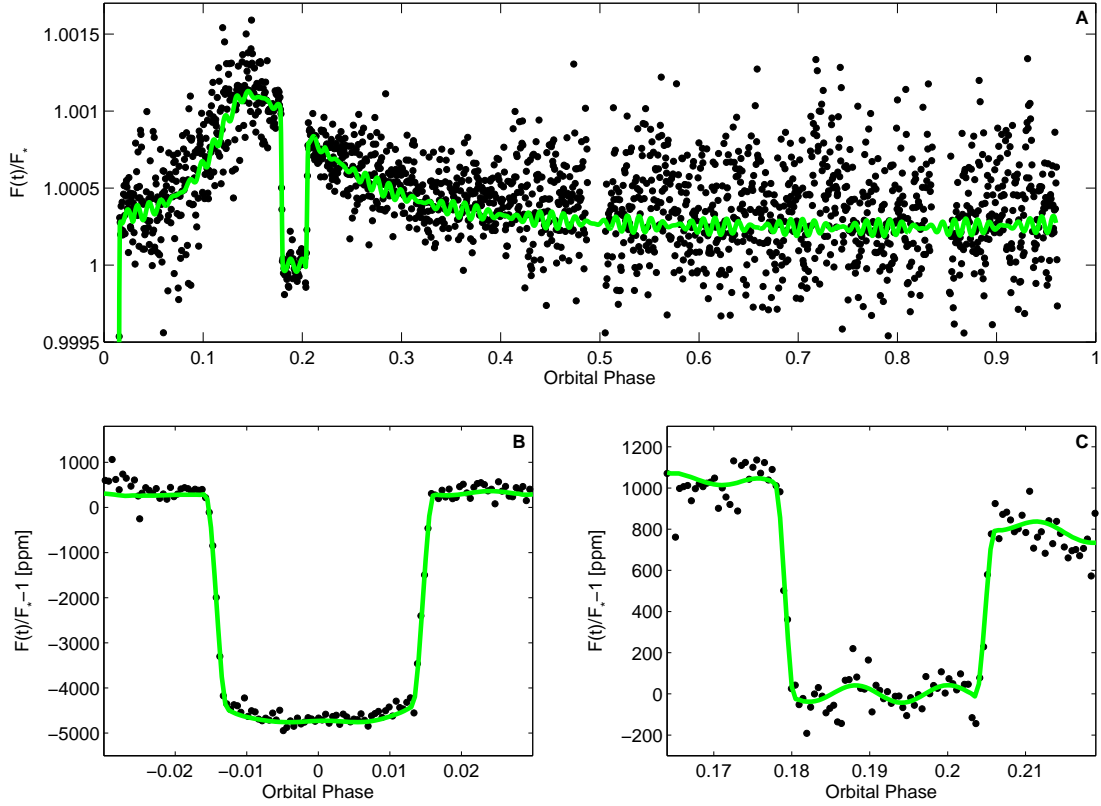}
  \vspace{-0.0cm}
  \caption{HAT-P-2\,b's photometry observed at 4.5 $\mu$m.  The photometry (black filled circles) combines $\sim350$ hours of observations obtained between 2011 July 9 and November 2015 and includes three primary and eighteen secondary eclipses. The photometry is presented normalized, with instrumental effects removed, phase-folded, and binned into intervals of 0.00025 in orbital phase--which corresponds to two minute intervals. The decrease in datapoint spread around occultation and periastron passage is due to the multiple observations at these orbital phases. We overplot the best-fit model light curve in green. \textbf{(A)} Phase-curve. \textbf{(B)} Transit. \textbf{(C)} Occultation.}
  \label{fig:photometry}

\end{figure*}

Our analysis also reveals distinct pulsations with a period of approximately 87 minutes in HAT-P-2's photometry (Figure\,\ref{fig:photometry}). The oscillations are observed during the occultations, which implies that if they are astrophysical in nature they must originate from the host star and not the planet. We verify that the pulsations are not due to a strong instrumental signal in a subset of the AORs which would then be diluted by the inclusion of additional observations without this signal. We first explore this possibility by showing that the oscillations are consistently visible across subsets of AORs. To do so, we randomly picked off the eighteen occultation AORs 25 random sets of 9 AORs, 25 random sets of 5 AORs, and 25 random sets of 3 AORs and included in our model a sine function. We found that all but two of the sets of 3 AORs and one of 5 AORs yield a consistent detection of the pulsation. The consistency of the retrieved pulsation implies that, although the pulsation is not detectable from an individual AOR, the SNR of 5 AORs combined is sufficient to yield a $>2\sigma$ detection. Furthermore, its consistency over these different AOR subsets indicates its coherence. We explore this further by including a sine curve function in the model and allowing its amplitude to vary individually for each occultation AOR. We find that the individual amplitudes scatter around 40 p.p.m., while the significance of the pulsation detection in each individual visit is less than 1 $\sigma$ (Figure\,\ref{fig:ind_eclipse_depths_and_oscillation}.(B)). The pulsations are hence present in each individual occultation AORs but its study requires fitting multiple AORs.

\begin{figure*}[!ht]
  \vspace{-0.0cm}\centering\includegraphics[trim=0mm 0mm 0mm 00mm, clip=true,width=17cm,height=!]{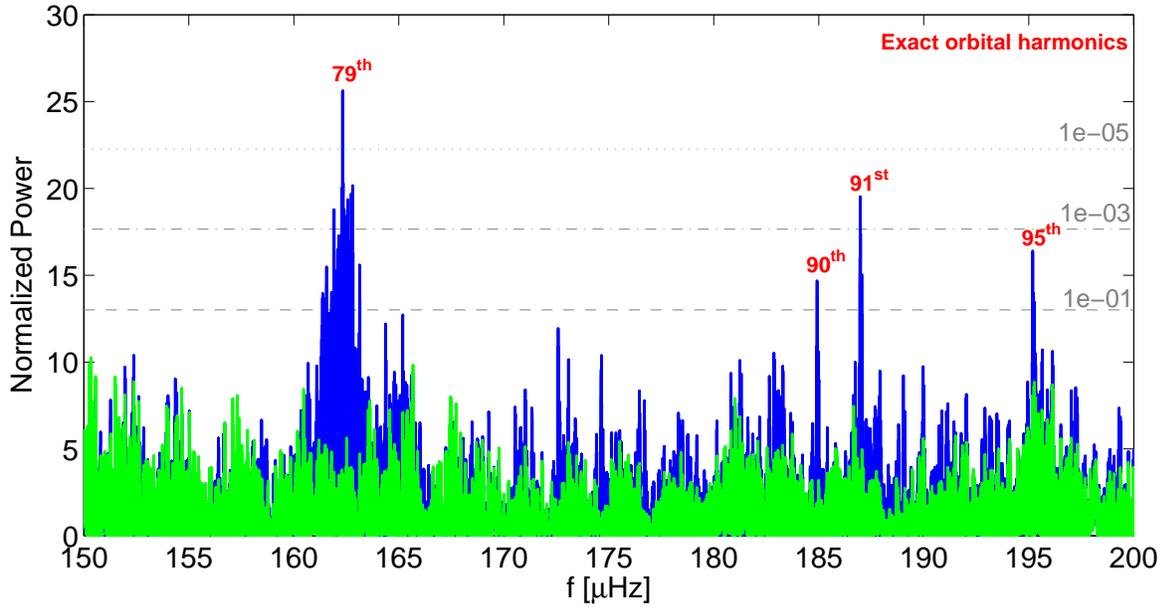}
  \vspace{-0.0cm}
  \caption{Periodogram of the 4.5 micron photometry after subtracting the best-fit model. The periodogram prior to accounting for the pulsations is shown in blue and the periodogram of the residuals after the best-fit pulsation model has been removed is shown in green. Significance levels are expressed in terms of false positive probability via horizontal grey lines.}
  \label{fig:perio}

\end{figure*}

\begin{figure*}[!ht]
  \vspace{-0.0cm}\centering\includegraphics[trim=0mm 0mm 0mm 00mm, clip=true,width=17cm,height=!]{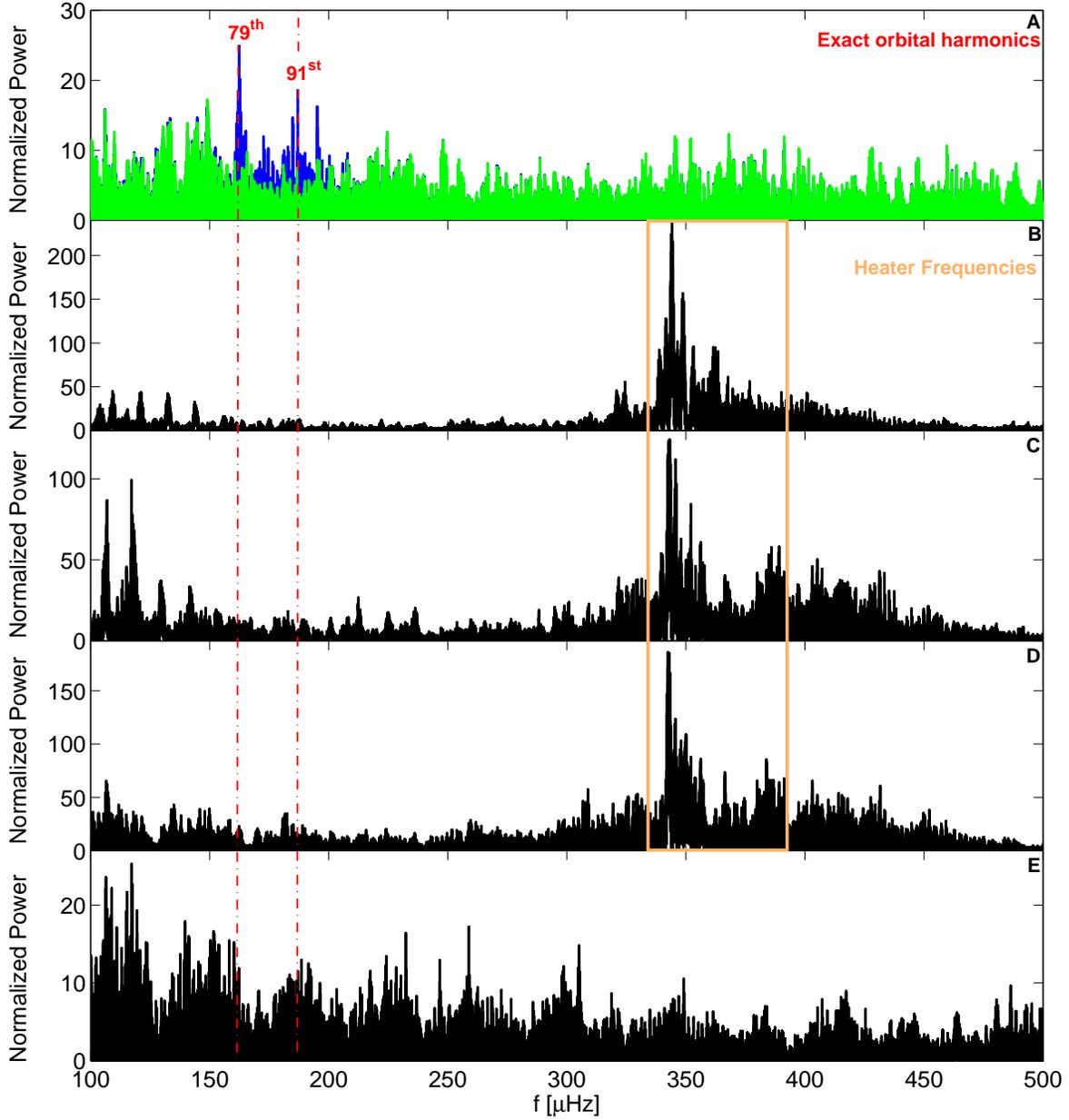}
  \vspace{-0.0cm}
  \caption{Periodogram of the 4.5 micron photometry compared to the periodograms of the point spread function (PSF) change in position along x/y \textbf{(B)/(C)}, the noise pixel parameter (proxy for the PSF shape) \textbf{(D)}, and the background contribution \textbf{(E)}. No energy is found in at the frequency corresponding to the pulsation detected in HAT-P-2's photometry ruling out their instrumental origin.}
  \label{fig:perio_comp}

\end{figure*}

The transit photometry, however, shows no sign of such short-period oscillations (Figure\,\ref{fig:photometry}.(B) and \textbf{Online Table}). Similar individual fits lead to an oscillation amplitude of 3$\pm$15 ppm in transit. The pulsations build coherently when we combine eighteen distinct epochs of data, which implies that the pulsation is a harmonic of the planet orbital frequency, as shown by the periodogram of the residuals (Figure\,\ref{fig:perio}).  The two most significant peaks correspond to HAT-P-2 b's 79$^{th}$ and 91$^{st}$ orbital harmonics, which are found to interfer constructively around occultations but destructively around transits as a result of the relative phase of the two harmonics. The pulsations are therefore astrophysical in nature unless the instrumental systematics appear to be harmonics of HAT-P-2 b's orbital frequency phased in such a way that they build coherently over all the different epochs. This scenario would require the modulation of the detector gain/noise or the PSF position/shape\footnote{The PSF shape can be affected by structural changes of the instrument (e.g., due to heating).} follows a repeated pattern of oscillatory changes that are not only harmonics of the planet's orbital frequency, but also constructively phased. We use the noise pixel parameter \cite{Mighell2005,Lewis2013} as a proxy for any changes in the PSF position and shape. We use the ``sky'' background as a proxy for gain variations and electronic noise. We find that the periodograms of the PSF position (Figures\,\ref{fig:perio_comp}(B) and (C)) and of the noise-pixel parameter (Figure\,\ref{fig:perio_comp}(D)) show no significant peak in the period range related to the pulsations while showing a significant signal at \textit{Spitzer}'s pointing oscillation period. Similarly, the periodogram of the ``sky'' background shows no specific peak around the 79$^{th}$ and 91$^{st}$ planetary harmonics as revealed in the photometry, ruling out the scenario of constructively-phased and adequately-modulated instrumental systematics. 

We therefore conclude that the pulsations are stellar and not instrumental in nature, and we account for them by including two sine functions in the final version of our global fits. We find that the pulsations have respective amplitudes and frequencies of 35$\pm$7 and 28$\pm$6ppm and 162.335$\pm$0.015 and 186.976$\pm$0.013 $\mu$Hz---respectively 79.006$\pm$0.007 and 91.006$\pm$0.007 times the orbital frequency.

\subsection{HAT-P-2's Radial Velocity}

\begin{figure*}
  \centering\includegraphics[trim=0mm 00mm 0mm 00mm, clip=true,width=18cm,height=!]{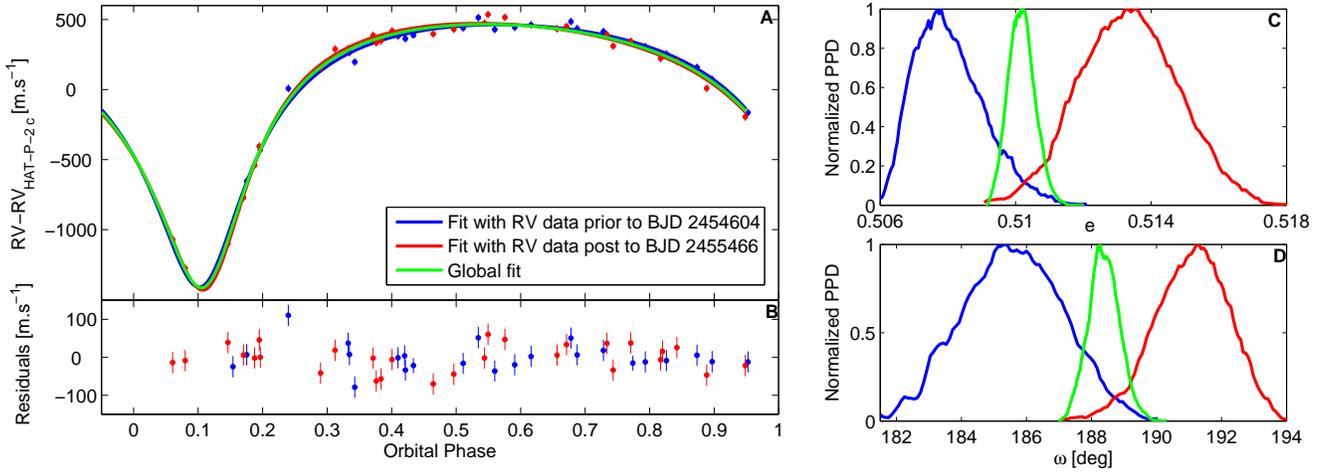}
  \vspace{-0.0cm}
  \caption{Complementary insights from HAT-P-2's radial velocity measurements. \textbf{(A)} HAT-P-2's RV measurement folded over HAT-P-2\,b's period with long-term variation in the RV signal due to substellar object c removed. Measurements are shown as filled circled with their error bars and best fits as solid lines. Objects colored in green, blue, and red correspond respectively to quantities related to our global analysis, analysis using RV data prior to BJD 2454604, and analysis using RV data prior to BJD 2454604. \textbf{(B)} Residuals of the RV fits. \textbf{(C)} Normalized posterior probability distribution (PPD) of HAT-P-2\,b's orbital eccentricity. \textbf{(D)} Normalized PPD of HAT-P-2\,b's periastron argument. The RV measurements show evidence of HAT-P-2\,b's evolution at the 4.5-$\sigma$ level on $e$ and $\omega$, which is inconsistent with the insight gained from the ultra-precise eclipse times derived from \textit{Spitzer}'s photometry.}
  \label{fig:RV}

\end{figure*}

HAT-P-2's radial velocity (RV) measurements provide a complementary perspective to \textit{Spitzer}'s. In addition to showing a high level of stellar jitter ($\sim 36$m/s, discussed in Section\,\ref{sec:dis_rv}), the RV measurements suggest that HAT-P-2\,b's orbit is evolving in a way that is inconsistent with our previous fits to the photometry alone. Figure\,\ref{fig:RV} shows that an independent analysis of the first half of HAT-P-2's RV data (obtained prior to BJD 2454604) yields values for $e$ and $\omega$ that are inconsistent at the 4.5-$\sigma$ level with those from a fit of the second half of the data (obtained after BJD 2455466). This translates into respective variations of $e$ and $\omega$ of 8.9$\pm$2.8E$^{-4}$ and 0.91$\pm$0.31$^{\circ}$ per year, which is inconsistent with the limits derived from the primary and secondary eclipse times alone.

\section{Discussion}
\label{sec:dis}

\subsection{The Origin of HAT-P-2's Pulsations}
\label{sec:dis_puls}

The exact correspondence of the stellar pulsation frequencies with harmonics of the planet's orbital frequency implies a connection between the two. Two scenarios are possible; either the pulsations are altering the planet's orbit or the planet is causing the pulsation. Mode-planet gravitational interactions could synchronize the planet's orbital frequency to an integer harmonic of a mode oscillation period, analogous to two weakly coupled oscillators that synchronize in the presence of weak damping. However, the detection of multiple harmonics of the planet's orbital frequency rules out this scenario. 

We next consider the hypothesis that the observed stellar pulsations are induced by the orbiting planet. Tidally excited pulsations have been previously observed in several eccentric stellar binaries \citep{Welsh2011,Thompson2012,Beck2014} known as ``heartbeat'' stars. The characteristic feature of a tidally excited pulsation is that its frequency is an exact integer harmonic of the orbital frequency, because it arises from a resonantly forced stellar pulsation mode \citep{Derekas2011,Fuller2012,Burkart2012,Oleary2014,Smullen2015}.

In order to understand the nature of the detected pulsations, we calculate the stellar response to tidal forcing. We first calculate the non-adiabatic stellar pulsation mode spectrum of HAT-P-2 using the MESA evolution code \citep{Paxton2011,Paxton2013} to describe the stellar evolution and the GYRE pulsation code \citep{Townsend2013} to compute the pulsation modes. We then calculate the response of each pulsation mode to the tidal forcing, and compute the amplitude and frequencies of each tidally excited pulsation. Our models indicate that although HAT-P-2\,b can in principle create luminosity fluctuations of this amplitude in its host, we expect these pulsations to have much lower frequencies, in the vicinity of twice the periastron orbital frequency \citep{Hut1981}---approximately $10~\mu$Hz (period of approximately 28 hrs).  Our models are unable to reproduce a tidally excited pulsation at the 79$^{th}$ or 91$^{th}$ orbital harmonics at the observed amplitude; the strength of the forcing at these harmonics is many orders of magnitude weaker than at the tidal forcing peak near the 6$^{th}$ orbital harmonic.

Although HAT-P-2 lies near the extreme cool ends of the $\gamma$-Doradus and $\delta$-Scuti instability strips \citep[$T_{eff} \sim $ 6300K, log$_{10}$g$\sim$4.16, $R \sim$ 1.54 $R_{Sun}$,][]{Uytterhoeven2011,Balona2015}, it shows no evidence for the large amplitude ($\sim$mmag) pulsations typically observed in these pulsators. Nonetheless, it is possible that the observed pulsations correspond to low-amplitude $\delta$-Scuti pulsation modes, as our non-adiabatic pulsation calculations indicate that $l\!=\!1$ and $l\!=\!2$ pulsation modes are unstable near the observed frequencies of $f \! \sim$175 $\mu$Hz, although we caution that mode growth rates--the parameter indicating stability of a pulsation mode--may be altered by a more realistic treatment of convective flux perturbations. In our models, the unstable modes correspond to low-order ($n \! < \! 3$) mixed pressure-gravity modes. In principle, these modes can couple strongly with HAT-P-2\,b because they produce relatively large gravitational perturbations (i.e., they are similar to fundamental modes). However, our models suggest that they have frequencies that are too large to resonantly interact with the planet.

The failure of our models to explain the observed high frequency pulsations suggests that current tidal theories may be incomplete. One tentative possibility is that non-linear effects are important. The difference between the observed pulsations is $\simeq$12 times the orbital frequency where resonant interactions with the planet can occur. This sort of non-linear mode splitting at orbital harmonics has been observed in stellar systems \citep[see e.g.,][]{Hambleton2013} and could occur in the HAT-P-2 system as well.

\subsection{Tidally-Induced Pulsations and RV Measurements}
\label{sec:dis_rv}

Stellar pulsations cause RV variations of the stellar photosphere \citep[e.g.,][]{Willems2002,Welsh2011}. The RV amplitude of the photospheric motion due to a pressure mode is $\Delta v \sim 2 \pi f \Delta R$, where $\Delta R$ is the mode surface displacement. Our non-adiabatic pulsation calculations indicate that the stellar surface displacements and luminosity variations are approximately related by $\Delta R/R \sim 2 \Delta L/L$ for modes near the observed frequencies. The observed photometric amplitudes imply $\Delta v \sim 60 \, {\rm m/s}$, consistent with HAT-P-2's RV jitter. Mode identification would be required for a more precise calculation, but we posit that HAT-P-2's RV jitter and HAT-P-2\,b's apparent orbital evolution in the RV data is mainly produced by the pulsations. 

\section{Conclusion}

Our study provides a new perspective on the planet-star interactions in HAT-P-2's system. The extensive coverage of the system at 4.5 $\mu$m allows us to better disentangle the instrumental systematic from the planet's transient heating, reconciling observations and atmospheric models. The photometric observations also show no sign of orbit-to-orbit variability nor of orbital evolution but reveals high-frequency low-amplitude stellar pulsations that correspond to harmonics of the planet's orbital frequency, supporting their tidal origin. Current stellar models are however unable to reproduce these pulsations. HAT-P-2's RV measurements exhibit a high level of jitter and support the orbital evolution of HAT-P-2\,b inconsistent with the ultra-precise eclipse times. The inability of current stellar models to reproduce the observed pulsations and the exotic behavior of HAT-P-2's RV indicate that additional observations and theoretical developments are required to understand the processes at play in this system.

Future missions such as \textit{TESS}, \textit{PLATO}, and \textit{CHEOPS} will provide further insights into the nature of HAT-P-2's pulsations and its interaction with its eccentric companion. The detection of multiple unexpected high-frequency pulsations as HAT-P-2's could be used to indirectly hint at the presence of a companion--regardless of their orbital inclination, to be later confirmed with traditional techniques such as direct imaging or RV.

\acknowledgments

This work is based on observations made with the Spitzer Space Telescope, which is operated by the Jet Propulsion Laboratory, California Institute of Technology, under contract to NASA. Support for this work was provided by JPL/Caltech. J.d.W. was further supported by the WBI (Wallonie-Bruxelles International) under the WBI-World Excellence Fellowship Program. A.S. performed this work in part under contract with the California Institute of Technology (Caltech) funded by NASA through the Sagan Fellowship Program executed by the NASA Exoplanet Science Institute. V.A. is funded by the Stellar Astrophysics Centre via the Danish National Research Foundation (Grant DNRF106). The research was supported by the ASTERISK project (ASTERoseismic Investigations with SONG and Kepler) funded by the European Research Council (Grant agreement no.: 267864). Radial velocity data presented herein were obtained at the W. M. Keck Observatory using time granted by the California Institute of Technology, UC Berkeley, and the University of Hawaii. We thank the observers who contributed to the measurements reported here and acknowledge the efforts of the Keck Observatory staff. We extend special thanks to those of Hawaiian ancestry on whose sacred mountain of Mauna Kea we are privileged to be guests. 

J.d.W. thanks T. Rogers, V. Stamenkovi{\'c}, A. Zsom, B.-O. Demory,  M. Gillon, S. Seager, and V. Van Grootel for useful discussions during the preparation of this manuscript.

\bibliographystyle{Biblio/apj}


\clearpage

\section{Online Table}

\begin{table*}[!b]
	\vspace*{-0cm}\caption{Description of HAT-P-2's AOR$\mathbf{^{\mathrm{a}}}$ at 4.5 $\mu$m}
	\label{tab:AOR}
		
	\setlength{\extrarowheight}{-10pt}

	\hspace*{-1cm}\scriptsize{\begin{tabular}{cc|ccc|cc}
		\hline\hline
		\textbf{AORKEY($\mathbf{\bigoplus^{\mathrm{b}}}$)} & \textbf{Start time [UT]} & Eclipse time [BJD-2455000] & Eclipse depth [p.p.m] & Pulsation amplitude [p.p.m.] & \textbf{Aperture$\mathbf{^{\mathrm{c}}}$ [px]}  & \textbf{$\beta_{red}$ $\mathbf{^{\mathrm{d}}}$}\\

		\hline
42789632(O) & 7/9/2011 1:51	& 5751.8763$\pm$0.0012 &	1120$\pm$83 & 36$\pm$90 & $\sqrt{\tilde{\beta}}$-0.6 (2.1) &	1.55\tabularnewline
42789888(P)  &	7/10/2011 1:36	 &  &  & & $\sqrt{\tilde{\beta}}$-0.6	(1.9) &	1.4\tabularnewline
43962624(P)	& 7/11/2011 3:16 &  &  &  & $\sqrt{\tilde{\beta}}$-0.6	(1.9) &	1.2\tabularnewline
43962880(P)	& 7/12/2011 3:01 &  &  &  & $\sqrt{\tilde{\beta}}$-0.6 (2.1) &	1.5\tabularnewline
43963136(T)	& 7/13/2011 2:12 & 5756.42696$\pm$0.00047 &	4961$\pm$72 & 12$\pm$42&  $\sqrt{\tilde{\beta}}$-0.6 (1.9) &	1.4\tabularnewline
43963392(O)	& 7/14/2011 1:56 &	5757.5081$\pm$0.0013 &	1116$\pm$117	& 32$\pm$83 &  $\sqrt{\tilde{\beta}}$-0.6 (1.8)& 1.45\tabularnewline
46473216(O)	& 10/31/2013 4:49&	6394.0922$\pm$0.0014 & 839$\pm$93 & -56$\pm$56&$\sqrt{\tilde{\beta}}$-0.5 (1.7)&	1.15\tabularnewline
46473472(O)	& 10/19/2013 22:45 & 6422.2591$\pm$0.0016 &	858$\pm$97 &76$\pm$46&$\sqrt{\tilde{\beta}}$-0.5 (1.6) &	1.6\tabularnewline
46473728(O)	& 10/14/2013 7:21 & 6427.8928$\pm$0.0011 & 854$\pm$83&78$\pm$45&$\sqrt{\tilde{\beta}}$-0.5	(1.5) &	1.05\tabularnewline
46474496(O)	& 9/21/2013 18:24 &	6489.8599$\pm$0.0013 &	887$\pm$96&	-16$\pm$45&$\sqrt{\tilde{\beta}}$-0.4 (2.1) &1.55\tabularnewline
46474752(O)	& 9/16/2013 3:11 & 6495.4952$\pm$0.0015 &899$\pm$110 & 39$\pm$49 &	$\sqrt{\tilde{\beta}}$-0.6 (1.8) &	1.2\tabularnewline
46475008(O) & 8/30/2013 5:55 & 6529.2941$\pm$0.0014 & 1002$\pm$96 &99$\pm$42 & $\sqrt{\tilde{\beta}}$-0.5	(1.55) & 1.3\tabularnewline
46475264(O) & 8/24/2013 14:26 & 6534.9295$\pm$0.0013 & 959$\pm$86 & 20$\pm$41 & $\sqrt{\tilde{\beta}}$-0.6	(1.45) & 1.4\tabularnewline
46475520(O)	& 7/21/2013 19:19 & 6540.5602$\pm$0.0010 & 802$\pm$97 & 96$\pm$49 & $\sqrt{\tilde{\beta}}$-0.7	(1.6) &	1.3\tabularnewline
46475776(O)	& 7/16/2013 4:21 & 6551.8272$\pm$0.0012 & 974$\pm$84 & 7$\pm$43 &	$\sqrt{\tilde{\beta}}$-0.5 (1.5) &	1.15\tabularnewline
46476032(O)	& 5/15/2013 4:55 & 6557.4625$\pm$0.0011 & 987$\pm$90 & 37$\pm$41 &	$\sqrt{\tilde{\beta}}$-0.6	(1.45) & 1.2\tabularnewline
46476288(O)	& 5/9/2013 13:53 & 6579.9946$\pm$0.0010 &	901$\pm$88 & 69$\pm$39 & $\sqrt{\tilde{\beta}}$-0.5	(1.55) & 1.45\tabularnewline
46476544(O)	& 4/11/2013 9:42 & 6585.6294$\pm$0.0013 & 750$\pm$83 & 18$\pm$48 & $\sqrt{\tilde{\beta}}$-0.5	(1.55) & 1.35\tabularnewline
46477312(P)	& 9/5/2013 21:46 &  &  &  & $\sqrt{\tilde{\beta}}$-0.6 (1.55) &	1.6\tabularnewline
46477568(P)	& 9/5/2013 9:46	 &  &  &  & $\sqrt{\tilde{\beta}}$-0.6	(1.45) & 1.3\tabularnewline
46477824(O) & 9/4/2013 21:45 & 6591.2611$\pm$0.0010 &890$\pm$80 & 94$\pm$40 & $\sqrt{\tilde{\beta}}$-0.5 (1.55) & 1.2\tabularnewline
46478336(P)	& 10/26/2013 14:45	 &  &  &  & $\sqrt{\tilde{\beta}}$-0.5	(1.5) &	1.2\tabularnewline
46478592(P)	& 10/26/2013 2:45	 &  &  &  & $\sqrt{\tilde{\beta}}$-0.4	(1.6) &	1.2\tabularnewline
46478848(O)	& 10/25/2013 14:44 & 6596.8941$\pm$0.0012 &	918$\pm$76 & -20$\pm$38 & $\sqrt{\tilde{\beta}}$-0.5 (1.55) & 1.4\tabularnewline
57787136(T)	& 10/21/2015 13:20 & 7316.89688$\pm$0.00047 & 4923$\pm$66 & -40$\pm$74 & $\sqrt{\tilde{\beta}}$-0.7	(1.35) & 1.1\tabularnewline
57786880(T) & 11/18/2015 17:39 & 7345.06511$\pm$0.00051 & 4948$\pm$71 & 15$\pm$71 & $\sqrt{\tilde{\beta}}$-0.7 (1.4) & 1.05\tabularnewline
57787648(O)	& 11/19/2015 19:28 & 7346.1474$\pm$0.0012 & 102$\pm$99 & 19$\pm$68 & $\sqrt{\tilde{\beta}}$-0.5	(1.6) &	1.45\tabularnewline
57787392(O)	& 11/25/2015 10:49 & 7351.7813$\pm$0.0011 &	864$\pm$87 & 33$\pm$85 & $\sqrt{\tilde{\beta}}$-0.6	(1.4) &	1.35\\

\end{tabular}}	
 
\begin{list}{}{}
\item[$^{\mathrm{a}}$] {AORs are composed of data sets (FITS files), each data set corresponds to 64 individual subarray images of 32x32 px. For the present AORs, the exposure time was set to 0.4 sec.}
\item[$^{\mathrm{b}}$] {AORKEY target: T, O, or P respectively for transit, occultation, or phase curve measurements.}
\item[$^{\mathrm{c}}$] {We optimize our choice of aperture individually for each AOR. We find that time-varying apertures equal to the square root of the noise pixel parameter $\tilde{\beta}$ best reduce the scatter in the final time series. The average aperture radii are shown in parentheses.}
\item[$^{\mathrm{d}}$] {The $\beta_{red}$ coefficients are used to estimate the fraction of remaining time-correlated noise in our final time-series and to account for it in our error estimates. To do so, we follow a procedure similar to ref. \cite{Winn2008}.}

\end{list}

\end{table*}

\end{document}